# Review and Prospect: NMR Spectroscopy Denoising & Reconstruction with Low Rank Hankel Matrices and Tensors

Tianyu Qiu[a], Zi Wang[a], Huiting Liu[a], Di Guo[b], Xiaobo Qu*[a]

**Abstract:** Nuclear Magnetic Resonance (NMR) spectroscopy is an important analytical tool in chemistry, biology, and life science, but it suffers from relatively low sensitivity and long acquisition time. Thus, improving the apparent signal-to-noise ratio and accelerating data acquisition become indispensable. In this review, we summarize the recent progress on low rank Hankel matrix and tensor methods, that exploit the exponential property of free induction decay signals, to enable effective denoising and spectra reconstruction. We also outline future developments that are likely to make NMR spectroscopy a far more powerful technique.

## 1. Introduction

### 1.1. Overview of NMR Spectroscopy

Nuclear Magnetic Resonance (NMR) spectroscopy plays an irreplaceable role in the science of chemistry, biology, and medicine, such as structure detection [1], quantitative analysis [2], reaction monitoring [3], disease diagnosis [4], and other fields.

Relatively low sensitivity of the spectra is a fundamental problem that is recurrently addressed in the development of the NMR methodology. Discarding part of data points at the end of FID signals is an intuitive and effective way but at the cost of reducing the spectra resolution. Signal averaging is another common way to enhance the signal-to-noise ratio (SNR). However, this approach suffers from a long acquisition time. In order to improve the SNR efficiently, if taking both resolution and acquisition time into consideration, denoising becomes an important process in NMR data processing.

It is well known that multi-dimensional spectroscopy can provide fruitful chemical information and prevent the peaks overlapping of the macromolecule. As the duration of NMR experiments increases exponentially with spectral dimensionality, another critical problem in developing the NMR technique is the long acquisition time. Non-Uniform Sampling (NUS) [5] is widely used to accelerate the acquisition of experimental data by acquiring partial data, and it needs advanced reconstruction methods to obtain full spectra. To obtain high-quality spectra, it is possible to introduce proper characteristics of NMR signals as constraints to solve the reconstruction problem.

In this review, we review the state-of-the-art low rank Hankel matrix and tensor methods to address the problem of denoising and NUS spectra reconstruction. We believe these methods would be very valuable to the NMR community since they exploit the intrinsic exponential property of Free Induction Decay (FID) signals.

### 1.2. Mathematical Model and Its Hankel Matrix

Typically, the time domain signals of NMR, which are often called FID, can be modeled as the superposition of several decaying exponential signals [5e, 6, 9]. A standard 1D signal is of the form

$$x_n = \sum_{r=1}^{R} c_r e^{(i2\pi f_r - \tau_r)n\Delta t}, \quad (1)$$

Where $x_n$ denotes the $n^{th}$ element of vector $\mathbf{x} \in \mathbb{C}^{N \times 1}$. $c_r$, $f_r$, $\tau_r$ and $R$ denote the complex amplitudes, resonance frequencies, damped factors, and the number of exponentials, respectively. And $\Delta t$ is the sampling interval.

The process of rearranging a 1D FID signal $\mathbf{x}$ into a Hankel matrix has been illustrated in Figure 1(a) and (b). The Hankel matrix given by $\mathbf{x}$ is written as

$$\mathbf{H_x} = \begin{pmatrix} x_1 & x_2 & \cdots & x_Q \\ x_2 & x_3 & \cdots & x_{Q+1} \\ \vdots & \vdots & \ddots & \vdots \\ x_{N-Q+1} & x_{N-Q+2} & \cdots & x_N \end{pmatrix}. \quad (2)$$

Define the operator $\mathcal{R}: \mathbb{C}^{N \times 1} \to \mathbb{C}^{(N-Q+1) \times Q}$, and $Q$ is larger than the number of exponential components of $\mathbf{x}$. In general, the Hankel matrix is defined as a square matrix, i.e., $Q = (N+1)/2$ when $N$ is an odd number, so that the Hankel matrix is capable of presenting enough exponential components. The Hankel matrix is close to a square matrix, i.e., $Q = N/2$, when $N$ is an even number. Such that for any $\mathbf{x} \in \mathbb{C}^{N \times 1}$,

$$\mathbf{H_x} = \mathcal{R}\mathbf{x}. \quad (3)$$

For a Hankel matrix $\mathbf{H_x}$, its matrix rank, i.e., the number of non-zero singular values is equal to the number of exponential components [5e, 6a, 7]. According to the Fourier theory, the number of exponential components is also equal to the number of spectral peaks. Thus, the rank of the Hankel matrix and the number of peaks should be the same. This property has been explored in NMR signal processing for many years, such as linear prediction [6a, 25], parameter estimation [26], denoising [8], and reconstructing [6a, 6b, 9].

Similar to the 1D case, it is possible to model a 2D FID signal as a sum of damped exponentials as

$$X_{n_1,n_2} = \sum_{r=1}^{R} c_r d_r^{n_1} \omega_r^{n_2}, \quad (4)$$

where $\mathbf{X} \in \mathbb{C}^{N_1 \times N_2}$, $d_r = e^{(i2\pi f_{1,r} - \tau_{1,r})\Delta_1}$, $\omega_r = e^{(i2\pi f_{2,r} - \tau_{2,r})\Delta_2}$, and $i$ denotes the imaginary unit satisfying $i^2 = -1$.

The sketch map of the procedure when converting 2D FID to a block Hankel matrix is shown in Figure 2.

Also, we denote an operator $\mathcal{B}$ to represent the transformation procedure and define it as [10]

$$\mathbf{B_X} = \mathcal{B}\mathbf{X} = \begin{bmatrix} \mathbf{X}_1 & \mathbf{X}_2 & \cdots & \mathbf{X}_{N_1-k_1+1} \\ \mathbf{X}_2 & \mathbf{X}_3 & \cdots & \mathbf{X}_{N_1-k_1+2} \\ \vdots & \vdots & \ddots & \vdots \\ \mathbf{X}_{k_1} & \mathbf{X}_{k_1+1} & \cdots & \mathbf{X}_{N_1} \end{bmatrix}, \quad (5)$$

where $\mathbf{B_X} \in \mathbb{C}^{k_1 k_2 \times (N_1-k_1+1)(N_2-k_2+1)}$ denotes a block Hankel matrix

[a] Tianyu Qiu, Zi Wang, Huiting Liu, Prof. X. Qu[*]
Department of Electronic Science, Fujian Provincial Key Laboratory of Plasma and Magnetic Resonance, Xiamen University
P.O.Box 979, Xiamen 361005 (China)
*E-mail: quxiaobo@xmu.edu.cn
[b] Prof. D. Guo
School of Computer and Information Engineering, Xiamen University of Technology, Xiamen 361024, China

and $k_1$ is a parameter. Each submatrix $\mathbf{X}_{n_1}$ in $\mathbf{B}_\mathbf{X}$ is a Hankel matrix, satisfying

$$\mathbf{X}_{n_1} = \begin{bmatrix} x_{n_1,1} & x_{n_1,2} & \cdots & x_{n_1,N_2-k_2+1} \\ x_{n_1,2} & x_{n_1,3} & \cdots & x_{n_1,N_2-k_2+2} \\ \vdots & \vdots & \ddots & \vdots \\ x_{n_1,k_2} & x_{n_1,k_2+1} & \cdots & x_{n_1,N_2} \end{bmatrix}, \quad (6)$$

where $k_2$ denotes another parameter. Similarly, the matrix $\mathbf{B}_\mathbf{x}$ also possesses the same property as the matrix $\mathbf{H}_\mathbf{x}$, which has been formed in several works [10b, 11]. This property is derived from Singular Value Decomposition (SVD). From this decomposition, a lot of information, such as the number and the line shape of peaks, which can be well utilized in the denoising/reconstruction problem.

As an example, the signal in Figure 1(a) is used to illustrate the definition of SVD as follows

$$\mathbf{H}_\mathbf{x} = \mathbf{U}\mathbf{\Sigma}\mathbf{V}^H, \quad (7)$$

where $\mathbf{\Sigma} \in \mathbb{R}^{(N-Q+1)\times Q}$ denotes a rectangular diagonal matrix, whose entries are singular values, and $H$ denotes the conjugate transpose. $\mathbf{U} \in \mathbb{C}^{(N-Q+1)\times(N-Q+1)}$ and $\mathbf{V} \in \mathbb{C}^{Q\times Q}$ are unitary matrices.

Before discussing the subspace, let us define subspaces of the 1D FID signal $\mathbf{x}$ and the Hankel matrix $\mathbf{H}_\mathbf{x}$ first.

(1) Subspaces of $\mathbf{x}$: A FID signal $\mathbf{x} \in \mathbb{C}^{N\times 1}$ can be linearly combined as

$$\mathbf{x} = \sum_{r=1}^R a_r \mathbf{x}_r, \quad (8)$$

where $\mathbf{x}_1$, $\mathbf{x}_2$, $\cdots$, and $\mathbf{x}_R$ are subspaces of $\mathbf{x}$.

According to Eq. (1), the subspace set could be $\left\{\mathbf{x}_r \in \mathbb{C}^{N\times 1} \middle| x_{r,n} = e^{(i2\pi f_r - \tau_r)n\Delta t}\right\}_{r=1,2,\cdots,R}$. Here we denote this subspace set as $D = \{\mathbf{d}_r\}_{r=1,2,\cdots,R}$ and the FID signal is expressed as

$$\mathbf{x} = \sum_{r=1}^R c_r \mathbf{d}_r.$$

(2) Subspaces of $\mathbf{H}_\mathbf{x}$: A Hankel matrix $\mathbf{H}_\mathbf{x} \in \mathbb{C}^{(N-Q+1)\times Q}$ with rank R can be linearly combined by a set of rank-1 matrices $\{\mathbf{H}_r \in \mathbb{C}^{(N-Q+1)\times Q}\}_{r=1,2,\cdots,R}$ as

$$\mathbf{H}_\mathbf{x} = \sum_{r=1}^R a_r \mathbf{H}_r, \quad (9)$$

where $\mathbf{H}_1$, $\mathbf{H}_2$, $\cdots$, and $\mathbf{H}_R$ are subspaces of $\mathbf{H}_\mathbf{x}$.

According to the definition of SVD in Eq. (7), the subspace set could be $\{\mathbf{H}_r | \mathbf{H}_r = \mathbf{u}_r \mathbf{v}_r^H\}_{r=1,2,\cdots,R}$ and the Hankel matrix is written as $\mathbf{H}_\mathbf{x} = \sum_{r=1}^R \sigma_r \mathbf{u}_r \mathbf{v}_r^H$, where $\mathbf{u}_r$ and $\mathbf{v}_r$ denote the $r^{th}$ column of matrices $\mathbf{U}$ and $\mathbf{V}$ in Eq. (7).

For the FID signal $\mathbf{x} \in \mathbb{C}^{N\times 1}$ expressed in Eq. (1), its subspace set is not unique. One possible subspace set could be $D$, another possible set could be $\{\mathbf{x}_r = \mathcal{R}^*\mathbf{u}_r\mathbf{v}_r^H\}_{r=1,2,\cdots,R}$ (Figure 1(e)), where $\mathcal{R}^*$ denotes an operator which converts a matrix to a vector via averaging the sum of entries on anti-diagonals of the matrix.

Since $\{\mathbf{d}_r\}_{r=1,2,\cdots,R}$ and $\{\mathbf{x}_r\}_{r=1,2,\cdots,R}$ are capable of combining $\mathbf{x}$, both of them contain the information of line shape, such as the location and the width, of peaks (Figure 1(c) and (e)). Compared with $\{\mathbf{d}_r\}_{r=1,2,\cdots,R}$, the physical meaning of $\{\mathbf{x}_r\}_{r=1,2,\cdots,R}$ is still unclear. However, this set is relatively easier to obtain in the applications, because it is derived from sets $\{\mathbf{u}_r\}_{r=1,2,\cdots,R}$ and $\{\mathbf{v}_r\}_{r=1,2,\cdots,R}$.

In some realistic cases, the number of peaks is found to be much lower than that of data points. Compared to the length of FID signals, take a 1D signal as an example, if the number of peaks is small enough, usually $R \leq 0.1N$, then the Hankel matrix can be considered as "low rank". In this paper, we focus on the denoising and reconstruction of spectra under these situations.

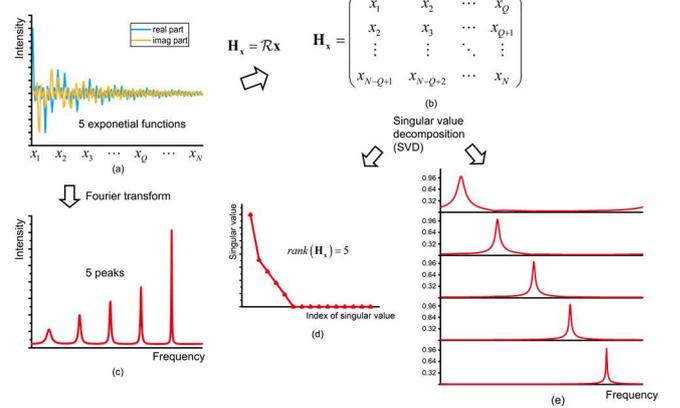

**Figure 1.** The relationship between a Hankel matrix converted by an FID signal and the number of exponential functions. (a) is a synthetic FID signal with 5 exponential functions; (b) denotes its Hankel matrix; (c) and (d) is the Fourier spectrum and the singular value spectrum of (a) and (b), respectively. (e) is subspaces of the signal in (a).

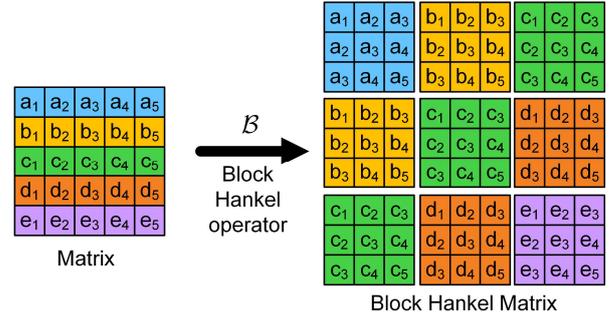

**Figure 2.** The procedure of converting 2D FID into a block Hankel matrix. Adapted from Ref. [11b].

### 1.3. What is Low Rank Hankel Matrix-based Denoising and Reconstruction?

During recent decades, this low rank Hankel property has been adopted in many powerful denoising [8] and reconstruction [6b, 9] methods. All these methods are based on solving the following problem

$$\min_\mathbf{x} \|\mathbf{y} - \mathcal{U}\mathbf{x}\|_2^2 \quad s.t. \ rank(\mathcal{R}\mathbf{x}) \leq R, \quad (10)$$

where $\mathcal{U}: \mathbb{C}^{N\times 1} \to \mathbb{C}^{M\times 1}$ denotes the sampling operator and $\|\cdot\|_2$ represents the $l_2$ norm. Due to the interference of noise, which is mostly described as Gaussian random distribution, the consistency with the measurement data is forced by minimizing the $l_2$ norm. Note that for $M = N$, we end up with a denoising problem. And if $M < N$, we are in the context of a reconstruction problem.

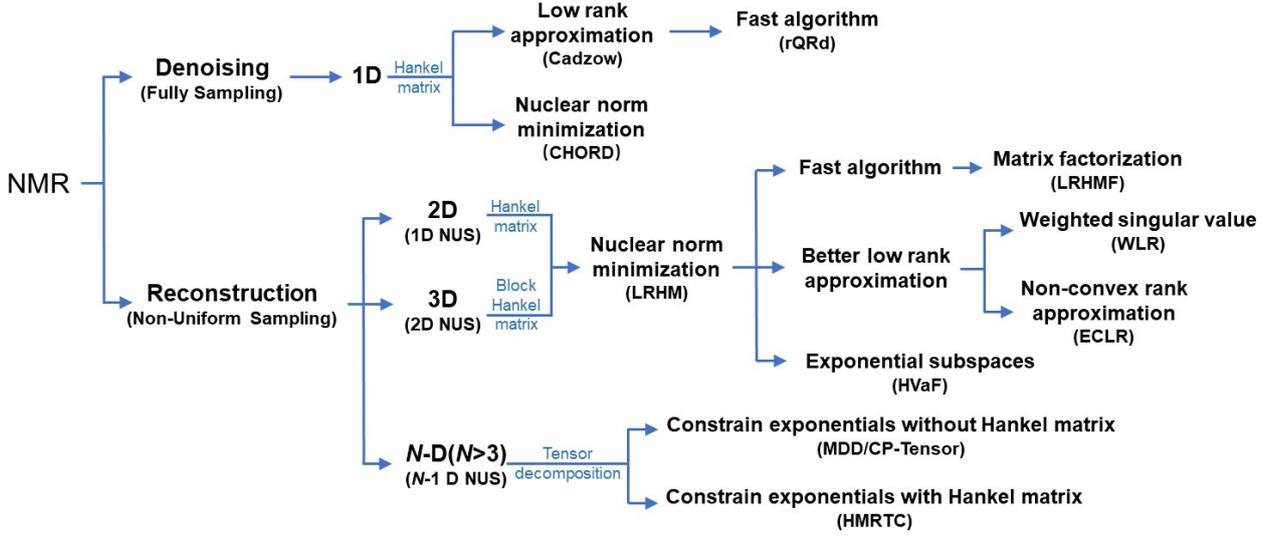

**Figure 3.** The organization of reviewed approaches in this paper.

**Table 1.** The parallel computational time and equipment for reviewed methods in this paper.

| Applications | Methods | Computational equipment | Data size | Computational time(second/hour) | Ref. |
|---|---|---|---|---|---|
| Denoising from fully sampled data | Cadzow | Intel i5-8259u ×1, 8G RAM | 1000 | 3.4s | [8a] |
| | CHORD | | | 2.2s | [8c] |
| | rQRd | | | 0.09s | [8b] |
| Reconstruction from Non-uniformly sampled data | LRHM | Intel E5-2650v4 (2.2GHz/12c) ×2, 128G RAM | 256×116 | 28.0s | [6b] |
| | | | 128×64×512 | 23.0h | [11b] |
| | LRHMF | | 256×116 | 17.8s | |
| | | | 128×64×512 | 0.34h | |
| | WLR | | 256×116 | 64.3s | [12] |
| | ECLR | | | 120.7s | [13] |
| | HVaF | | | 1.3h | [14] |

## 2. Low Rank Hankel Denoising

Numerous efforts have been made in the denoising of NMR spectroscopy by using the low rank property [8a, 8b, 15]. The denoising methods are divided into three types, full low rank decomposition, partial low rank decomposition, and convex low rank constraint. This section is dedicated to listing typical and the state-of-the-art methods in FID signals that exploit the low rank Hankel property and evaluate the performances of representative methods to analyze their advantages and shortcomings.

### 2.1. Full Low Rank Decomposition

Figure 1 implies that, for the FID signal $\mathbf{x}$ in Eq. (1) with the SVD in Eq. (7), the first $R$ singular values and the first $R$ columns of matrices $\mathbf{U}$ and $\mathbf{V}$ can represent the whole signal. For a measurement vector $\mathbf{y}$ corrupted by Gaussian noise $\mathbf{Z}$, its Hankel matrix tends to possess $R$ large singular values. However, the rank $R$ is difficult to determine in applications. Here we replace $R$ by $\hat{R}$ which denotes the estimated rank.

The simplest denoising method based on low rank Hankel matrix is SVD Truncation (TSVD), which holds the first $\hat{R}$ columns of matrices $\mathbf{\Sigma}$, $\mathbf{U}$ and $\mathbf{V}$ through the truncation. TSVD is based on solving the following problem

$$\min_{\mathbf{X}} \|\mathbf{Y}-\mathbf{X}\|_F^2 \ s.t. \ rank(\mathbf{X}) \leq \hat{R}, \quad (11)$$

where $\hat{R}$ denotes the estimate rank of $\mathbf{X}$.

However, in denoising, TSVD may cause the omission of peaks or the appearance of artifacts under relatively low SNR [8b].

Cadzow is another common tool in denoising of NMR spectroscopy [8a, 15a, 15c]. Compared with TSVD, this method constrains the truncated matrix owns Hankel structure. It is used to solve the following problem [8b]

$$\min_{\mathbf{x}} \|\mathbf{y}-\mathbf{x}\|_2^2 \ s.t. \ rank(\mathcal{R}\mathbf{x}) \leq \hat{R}. \quad (12)$$

Let us consider a map defined as $\mathcal{T}_{\hat{R}}: \mathbb{C}^{(N-Q+1)\times Q} \to \mathbb{C}^{(N-Q+1)\times Q}$. Let matrix $\mathbf{X} \in \mathbb{C}^{(N-Q+1)\times Q}$ with the SVD $\mathbf{X}=\mathbf{U}\mathbf{\Sigma}\mathbf{V}^H$, $\mathcal{T}_{\hat{R}}(\mathbf{X}) = \mathbf{U}_{\hat{R}}\mathbf{\Sigma}_{\hat{R}}\mathbf{V}_{\hat{R}}^H$, where matrices $\mathbf{U}_{\hat{R}}$, $\mathbf{\Sigma}_{\hat{R}}$ and $\mathbf{V}_{\hat{R}}$ are the first $\hat{R}$ columns of $\mathbf{U}$, $\mathbf{\Sigma}$, and $\mathbf{V}$, respectively. The operator $\mathcal{R}^*: \mathbb{C}^{(N-Q+1)\times Q} \to \mathbb{C}^{N\times 1}$ converts a Hankel matrix into a vector through summing the entries on the same anti-diagonal. Denote $\mathbf{w} \in \mathbb{R}^{N\times 1}$, and $w_k$ stands for the reciprocal of the number of the entries on $k^{th}$ anti-diagonal. $\circ$ is the Hadamard production.

For each step, the thresholding on the singular values of the Hankel matrix guarantees that $rank(\mathcal{R}\mathbf{x}_k) \leq \hat{R}$. But the thresholding may destroy the Hankel structure (Figure 4(c)). Thus, it is necessary to adopt an averaging on the sum of anti-diagonals of $\mathbf{H}_{k+1}$ ensures that $\mathbf{x}_{k+1}$ is given by a matrix that owns the Hankel matrix structure (Figure 4(d)).

**Table 2.** The solver of Cadzow.

**Initialization**: Input $\mathbf{y}$, $\hat{R}$, set iteration steps $t = 50$. And initialize $\mathbf{x}_0 = \mathbf{y}$.
**Main:**
**While:** $k < t$
    1) Update $\mathbf{H}_{k+1} = \mathcal{T}_{\hat{R}}(\mathcal{R}\mathbf{x}_k)$;
    2) Update $\mathbf{x}_{k+1} = \mathbf{w} \circ \mathcal{R}^*\mathbf{H}_{k+1}$.
**End while**
**Output:** $\mathbf{X}$.

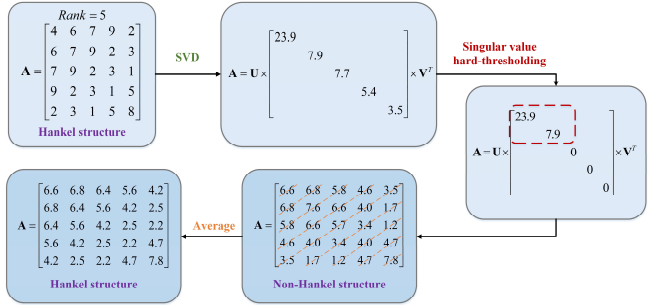

**Figure 4.** The sketch map for maintaining the Hankel matrix structure. (a) is a Hankel matrix. (b) denotes SVD of the matrix in (a). (c) stands for the thresholding on the singular values. (d) is the procedure of the averaging on the sum of anti-diagonals.

### 2.2. Partial Low Rank Decomposition

It is well known that SVD is a time-consuming procedure, particularly for large data size, which is common in the application of NMR spectroscopy. Some partial methods of low rank decomposition [8b, 15b] are proposed to accelerate the computation. These two methods, randomized SVD (rSVD) [15b] and randomized QR decomposition (rQRd) [8b, 15b], reduce the size of Hankel matrix by constructing a much smaller random matrix $\boldsymbol{\Omega}$ thereby decreasing the computation time effectively. Because of similar main concept, we only present the solver of rQRd as an example in Table 3.

rSVD and rQRd both reduce the computation time at the cost of retaining the noise residue and weakening the signal intensity, which results in the loss of low intensity peaks [15d].

To address this problem, Yu Yang et al. [15d] proposed a method that exploits a soft-thresholding generated from pre-denoised rSVD outcomes, which can remain the peak region and suppress noise region.

**Table 3.** The solver of rQRd.

**Initialization:** Input $\mathbf{y}$ and $\hat{R}$.
**Main:**
    1) Generate a random matrix $\boldsymbol{\Omega} \in \mathbb{C}^{Q \times \hat{R}}$;
    2) Form matrix $\mathbf{C} = (\mathcal{R}\mathbf{y})\boldsymbol{\Omega}$;
    3) Construct orthogonal matrix $\mathbf{Q} \in \mathbb{C}^{Q \times \hat{R}}$ such that $\mathbf{C} = \mathbf{QR}$ via QR factorization;
    4) Form matrix $\tilde{\mathbf{C}} = \mathbf{QQ}^*(\mathcal{R}\mathbf{y})$;
    5) Calculate $\mathbf{x} = \mathbf{w} \circ \mathcal{R}^* \tilde{\mathbf{C}}$.
**Output:** $\mathbf{X}$.

### 2.3. Convex Low Rank Hankel Constraint

The estimated rank $\hat{R}$, which is necessary in denoising methods above, is at least one parameter. Denoised performances of TSVD and Cadzow are highly dependent on this parameter. Although rSVD and rQRd have a more flexible $\hat{R}$, the dimension reduction of Hankel matrix, which is controlled by $\hat{R}$, still effects the denoised results.

There is a novel method, Convex Hankel lOw Rank Denoising (CHORD) [8c], which is expressed as

$$\min_{\mathbf{x}} \|\mathcal{R}\mathbf{x}\|_* + \frac{\lambda}{2}\|\mathbf{y} - \mathbf{x}\|_2^2, \quad (13)$$

where $\|\cdot\|_*$ denotes the nuclear norm, defined as the sum of the singular values. $\lambda$ is a regularization parameter balancing the nuclear norm and data consistency.

CHORD minimizes the rank of Hankel matrix, and uses the nuclear norm as a surrogate for the rank since the original issue is NP-Hard. Furthermore, it only has one parameter $\lambda$, and the parameter setting is automatic by theoretical guidance [8c].

Define the soft-thresholding operator which applies to $\mathbf{X} \in \mathbb{C}^{(N-Q+1)\times Q}$ as $\mathcal{S}_{1/\beta}(\mathbf{X}) = \mathbf{U}\,diag(\{\sigma_r - 1/\beta\}_+)\mathbf{V}^H$, where $diag(\cdot)$ is an operator which spans a vector into a diagonal matrix, and $t_+ = \max(0,t)$ [16]. Introducing variable $\mathbf{Z}$ and Lagrange multiplier $\mathbf{D}$, and utilizing Alternating Direction Method of Multipliers (ADMM) [17], the solution of Eq. (13) is given in Table 4.

**Table 4.** The solver of CHORD.

**Initialization**: Input $\mathbf{y}$, $\mathcal{R}$, set $\tau = \beta = 1$, and tolerance of solution in iterations $\eta = 10^{-8}$. Initialize the multiplier $\mathbf{D}_0 = \mathbf{1}$, the variable $\mathbf{Z}_0 = \mathbf{0}$, $\mathbf{x}_0 = \mathbf{y}$, and $\Delta\mathbf{x} = 10^{10}$.
**Main:**
**While:** $\Delta\mathbf{x} > \eta$
    1) Update
$$\mathbf{x}_{k+1} = (\lambda\mathbf{I} + \mathcal{R}^*\mathcal{R})^{-1}\left[\lambda\mathbf{y} + \mathcal{R}^*\left(\mathbf{Z}_k - \frac{\mathbf{D}_k}{\beta}\right)\right];$$
    2) Update $\mathbf{Z}_{k+1} = \mathcal{S}_{1/\beta}\left(\mathcal{R}\mathbf{x}_{k+1} + \frac{\mathbf{D}_k}{\beta}\right)$;
    3) Update $\mathbf{D}_{k+1} = \mathbf{D}_k + \tau(\mathcal{R}\mathbf{x}_{k+1} - \mathbf{Z}_{k+1})$;
    4) Compute $\Delta\mathbf{x} = \dfrac{\|\mathbf{x}_{k+1} - \mathbf{x}_k\|_2^2}{\|\mathbf{x}_{k+1}\|_2^2}$.
**End while**
**Output:** $\mathbf{X}$.

### 2.4. Conclusion on Low Rank Hankel Methods in Denoising

We analyze the denoised performances of a typical method, Cadzow, the state-of-the-art techniques, rQRd and CHORD on 1D $^1$H NMR data [8c]. This spectrum was acquired at 298 K on a Varian 500 MHz NMR system. The sample is a mixture of creatine, choline, magnesium citrate and calcium citrate. The concentration of these metabolites is 2:2:1:1.

All three methods produce reasonable and comparable denoised results under a high SNR circumstance (Figure 5(c)). For a relatively strong noise level (Figure 5(a)), Cadzow smooths the spectrum and offers a nice noise removing results. However, it also causes some peaks to be omitted (such as the peaks at 6.8 ppm). rQRd provides a spectrum with noticeable noise residue (orange lines in Figure 5(a)), which weakens low-intensity peaks (such as the peaks at 6.8 and 3.7 ppm). CHORD effectively removes noise and retains more details of peaks (Figure 5(a)).

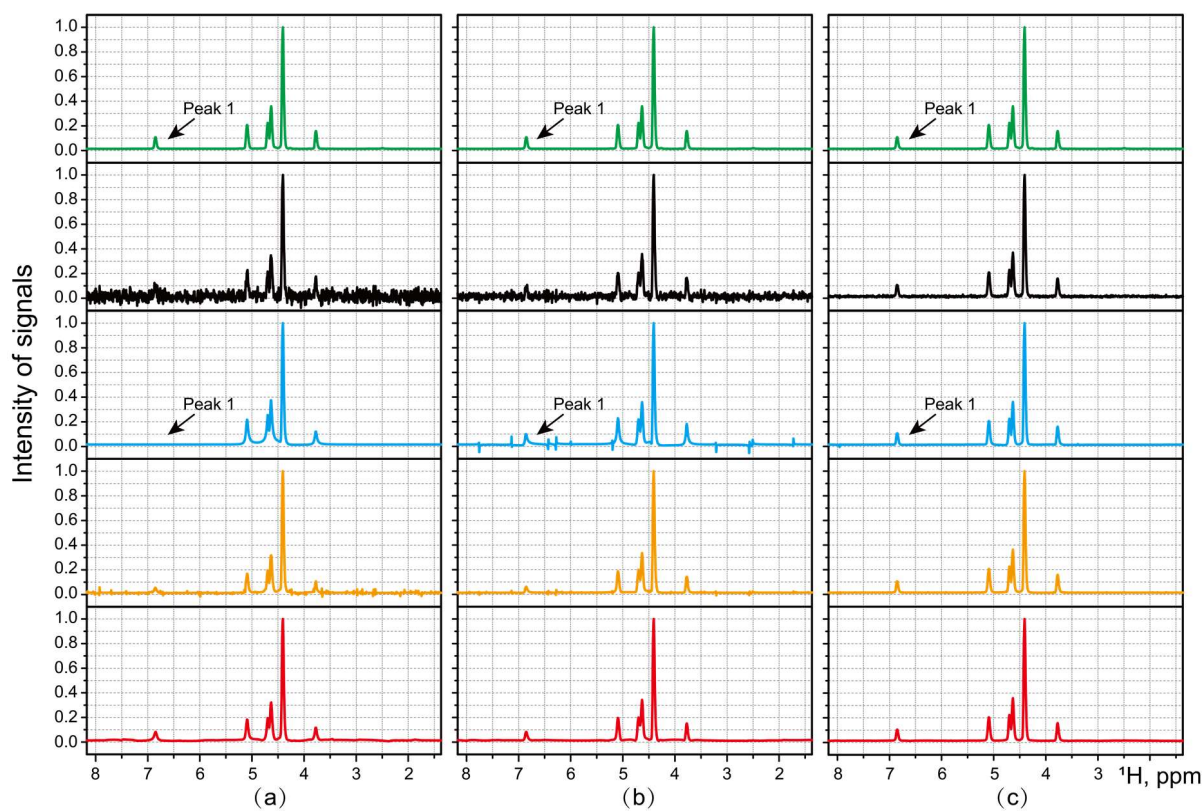

**Figure 5.** Denoised performances of a $^1$H spectrum of metabolites with $\sigma = 0.035$ (a), $0.020$ (b), and $0.005$ (c), respectively. The green lines denote the ground truth. The black lines indicate observations. Blue, orange, and red lines are denoised results of Cadzow, rQRd, and CHORD, respectively. Note: The results of Cadzow and rQRd that enable the lowest RLNE are presented here. Adapted from Ref. [8c].

## 3. Typical Low Rank Hankel Matrix/Tensor Reconstruction

To accelerate the acquisition of experimental data, a sampling schedule called NUS which circumvents the limitations of the Nyquist theorem, is widely used. In NUS, the location of sampling points can be completely random [5a, 5d], or satisfy certain distributions [5b, 5c]. One may also think of under-sampling and truncation as part of NUS.

Since the long acquisition time in high-dimensional spectroscopy is primarily due to the evolution, the NUS schedule is applied only in indirect dimensions. The sketch map (Figure 6) shows the procedure of random sampling in 2D and 3D spectroscopy. Solid circles stand for sampled data points, while hollow circles denote un-sampled data points. Accordingly, the reconstruction along the indirect dimension is sufficient to recover the entire spectrum. For example, a 2D NMR spectrum can be found by a series of 1D reconstructions. The region marked by a hollow red rectangle (or parallelogram) represents one reconstruction.

Previously, Linear Prediction (LP) [6a, 18, 25] in NMR data processing was one of the most typical techniques that exploit the low rank Hankel property, to reconstruct the spectra and estimate parameters. Recently, many novel methods utilizing the low-rankness are springing up. They all have good performance in reconstructing high-quality spectra from NUS data. We will discuss each method in more detail below.

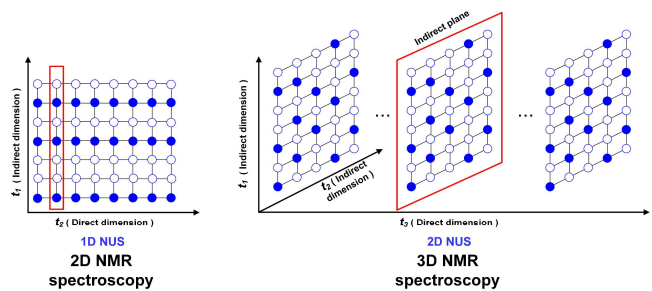

**Figure 6.** The sketch map of the random sampling schedule in the reconstruction of 2D and 3D NMR spectra. Adapted from Ref. [11b].

### 3.1. 2D and 3D Spectra Reconstruction with Low Rank Hankel Matrix

The Low Rank Hankel Matrix reconstruction method (LRHM) [6b] is a new scheme, which theoretically is well adapted to NMR spectroscopy [19]. The LRHM method can be formulated as a low rank matrix completion problem:

$$\min_{\mathbf{x}} \| \mathcal{R}\mathbf{x} \|_* + \frac{\lambda}{2} \| \mathbf{y} - \mathcal{U}\mathbf{x} \|_2^2 . \qquad (14)$$

In a nutshell, the high-quality spectrum can be reconstructed in two simple steps: 1) Convert the NUS FID signal into the Hankel matrix $\mathcal{R}\mathbf{x}$. 2) Solve $\mathbf{x}$ in Eq. (14) through ADMM.

Figure 7 shows the reconstruction of the $^1$H-$^{15}$N HSQC spectrum of the intrinsically disordered cytosolic domain of human CD79b protein from the B-cell receptor by LRHM, compared with the fully sampled reference spectrum. At a 35% sampling rate, LRHM offers a high-quality result. For the better analysis, 1D $^{15}$N traces of four representative peaks are presented, indicating that LRHM performs well at broad and weak peaks compared with Compressed Sensing (CS) [20].

**Table 5.** The solver of LRHM.

**Initialization**: Input: $\mathbf{y}, \mathcal{U}, \lambda, \beta = 1$, step size $\tau = \beta$, and tolerance of solution in iterations $\eta = 10^{-8}$. Initialize the dual variable $\mathbf{D} = \mathbf{1}$ and initialize $\mathbf{x} = \mathcal{U}^*\mathbf{y}$, $\mathbf{x}_{last} = \mathbf{x}$, $\Delta \mathbf{x} = 10^3$.
**Main:**
**While:** $\Delta \mathbf{x} \geq \eta$
1) Given $\mathbf{Z} = \mathcal{R}\mathbf{x}$, $\mathbf{D}$, update
$$\mathbf{x}_{k+1} = \left(\lambda \mathcal{U}^*\mathcal{U} + \beta \mathcal{R}^*\mathcal{R}\right)^{-1}\left[\lambda \mathcal{U}^*\mathbf{y} + \beta \mathcal{R}^*(\mathbf{Z}_k - \frac{\mathbf{D}_k}{\beta})\right]$$
2) Update $\mathbf{Z}_{k+1} = S_{\frac{1}{\beta}}\left(\mathcal{R}\mathbf{x}_{k+1} + \frac{\mathbf{D}_k}{\beta}\right)$
3) Update $\mathbf{D}_{k+1} = \mathbf{D}_k + \tau(\mathcal{R}\mathbf{x}_{k+1} - \mathbf{Z}_{k+1})$
4) Compute $\Delta \mathbf{x} = \|\mathbf{x}_{last} - \mathbf{x}\|/\|\mathbf{x}_{last}\|$

**Output:** $\mathbf{x}$.

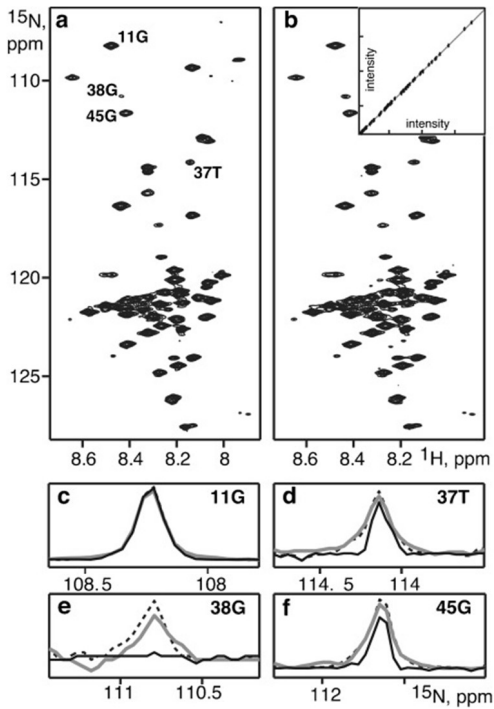

**Figure 7.** $^1$H-$^{15}$N HSQC spectrum of the cytosolic domain of CD79b. (a) fully sampling reference spectrum, (b) the LRHM reconstruction of 35% NUS data. (c)-(f) stand for the reconstruction of 4 peaks, 11G, 37T, 38G, and 45G, respectively. Dashed, gray, and black lines stand for the 1D $^{15}$N trace of four peaks in the reference, LRHM, and CS, respectively. Reproduced from Ref. [6b]. Copyright 2015, John Wiley and Sons.

In the reconstruction of 3D NMR spectra, the Block Hankel matrix is constructed (Figure 2) to replace the Hankel matrix. The LRHM for 3D NMR data is expressed as [11b]

$$\min_{\mathbf{X}} \|\mathcal{B}\mathbf{X}\|_* + \frac{\lambda}{2}\|\mathbf{Y} - \mathcal{U}\mathbf{X}\|_F^2, \quad (15)$$

where $\|\cdot\|_F$ denotes the Frobenius norm.

ADMM algorithm also can be utilized to solve Eq. (15). The solver is similar to that in the 2D case. Thus, it is not shown here.

### 3.2. Fast 2D and 3D Spectra Reconstruction with Low Rank Hankel Matrix

Ordinary methods of low rank Hankel matrix reconstruction include SVD, which consumes a lot of time in ADMM. It dramatically increases the reconstruction time, especially for solving the multi-dimensional spectral reconstruction problem.

An improved method called Low Rank Hankel Matrix Factorization (LRHMF) [11b] reduces the computation time dramatically by avoiding SVD.

Take the reconstruction of 2D NMR spectra as an example, LRHMF calculates $\|\mathcal{R}\mathbf{x}\|_*$ through solving

$$\min_{\mathbf{P},\mathbf{Q}} \frac{1}{2}\left(\|\mathbf{P}\|_F^2 + \|\mathbf{Q}\|_F^2\right), \quad s.t. \ \mathcal{R}\mathbf{x} = \mathbf{P}\mathbf{Q}^H. \quad (16)$$

The LRHMF method is based on solving the following problem:

$$\min_{\mathbf{P},\mathbf{Q},\mathbf{x}} \frac{1}{2}\left(\|\mathbf{P}\|_F^2 + \|\mathbf{Q}\|_F^2\right) + \frac{\lambda}{2}\|\mathbf{y} - \mathcal{U}\mathbf{x}\|_2^2 \quad s.t. \ \mathcal{R}\mathbf{x} = \mathbf{P}\mathbf{Q}^H. \quad (17)$$

The solver of LRHMF, which utilizes ADMM, is shown in Table 6.

**Table 6.** The solver of LRHMF.

**Initialization:** Input $\mathbf{y}$, $\mathcal{R}$, $\mathcal{U}$, set step size $\beta = 1$, convergence condition $\eta = 4 \times 10^{-2}$ and maximum number of iterations $K = 100$. Initialize the solution $\mathbf{x}_0 = \mathcal{U}^*\mathbf{y}$, the dual variable $\mathbf{D} = \mathbf{1}$. $\mathbf{P}_0$ and $\mathbf{Q}_0$ are complex random matrices with $(N-Q+1) \times 0.1N$ and $Q \times 0.1N$, respectively.
**Main:**
**While** ($\eta_k \geq \eta$) or ($k < K$), **do:**
1) Update
$$\mathbf{x}_{k+1} = \left(\lambda \mathcal{U}^*\mathcal{U} + \beta \mathcal{R}^*\mathcal{R}\right)^{-1}\left[\lambda \mathcal{U}^*\mathbf{y} + \beta \mathcal{R}^*\left(\mathbf{P}_k \mathbf{Q}_k^H - \frac{\mathbf{D}_k}{\beta}\right)\right];$$
2) Update $\mathbf{P}_{k+1} = \beta \left(\mathcal{R}\mathbf{x}_{k+1} + \frac{\mathbf{D}_k}{\beta}\right)\mathbf{Q}_k \left(\beta \mathbf{Q}_k^H \mathbf{Q}_k + \mathbf{I}\right)^{-1}$;
3) Update $\mathbf{Q}_{k+1} = \beta \left(\mathcal{R}\mathbf{x}_{k+1} + \frac{\mathbf{D}_k}{\beta}\right)^H \mathbf{P}_{k+1} \left(\beta \mathbf{P}_{k+1}^H \mathbf{P}_{k+1} + \mathbf{I}\right)^{-1}$;
4) Update $\mathbf{D}_{k+1} = \mathbf{D}_k + \left(\mathcal{R}\mathbf{x}_{k+1} - \mathbf{P}_{k+1}\mathbf{Q}_{k+1}^H\right)$;
5) Compute $\eta_k = \frac{\|\mathbf{x}_{k+1} - \mathbf{x}_k\|}{\|\mathbf{x}_k\|}$.

**End while**
**Output:** $\mathbf{x}$.

We compare the reconstruction results of LRHM and LRHMF on the $^1$H-$^{15}$N HSQC spectrum and analyze the performance through the correlation of peak intensities in Figure 8 and 9. Under the same computation time, LRHMF produces a spectrum with high-fidelity, while LRHM provides a spectrum with a large number of artifacts (Figure 8(b) and (d)). When the computation time is adequate for LRHM, reconstruction spectra of two methods are nice and comparable (Figure 8(c), (d), and Figure 9(a), (b)).

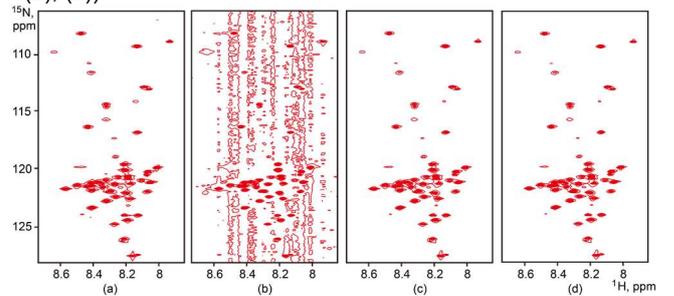

**Figure 8.** Reconstructed results of HSQC spectrum in Figure 7(a). (a) The fully sampled spectrum, (b) and (c) are reconstructed spectra at the time of 11.6s and 46.4s using LRHM, respectively, (d) the reconstructed spectrum at the time of 11.6s using LRHMF. Note: 35% data are acquired in NUS. The mentioned time is based on the reconstruction algorithms without parallel acceleration. Adapted from Ref. [11b].

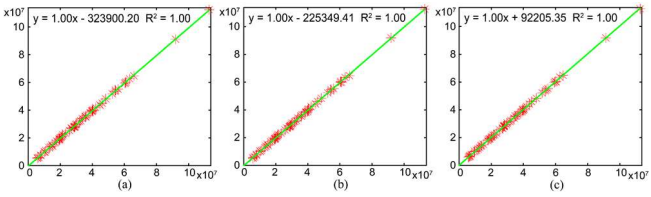

**Figure 9.** Peaks intensities correlation for the spectra in Figure 8. (a) and (b) are peak intensities correlations between the fully sampled spectrum and the reconstructions by LRHM and LRHMF, respectively; (c) is peak intensities correlation between two reconstructed spectra by LRHM and LRHMF, respectively. Note: The value of $R^2$ lies between 0 and 1 and $R^2$ is proportional to the correlation between the fully sampled and reconstructed spectrum. Reproduced from Ref. [11b]. Copyright 2017, IEEE.

In 3D case, similar to (15), LRHMF is expressed as

$$\min_{P,Q,X} \frac{1}{2}\left(\|\mathbf{P}\|_F^2 + \|\mathbf{Q}\|_F^2\right) + \frac{\lambda}{2}\|\mathbf{Y} - \mathcal{U}\mathbf{X}\|_F^2. \quad (18)$$

The performance of LRHMF has been tested on a 3D HNCO spectrum of U-[$^{15}$N, $^{13}$C] RNA recognition motifs domain of protein RNA binding motif protein 5 [21]. This spectrum was acquired on a Bruker AVANCE III 600 MHz spectrometer equipped with a cryogenic probe at 293K.

Compared with LRHM, the reconstruction of LRHMF is presented in Figure 10. Within 1.32h, LRHMF generates a satisfying spectrum with a high correlation of peak intensities, while LRHM provides a spectrum with a large amount of missing and weakened peaks.

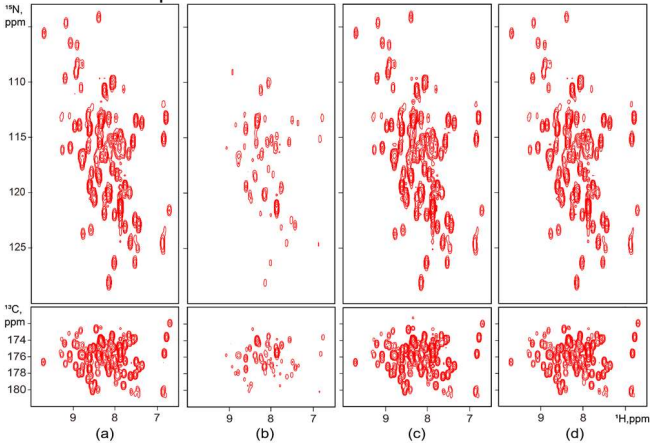

**Figure 10.** The $^1$H-$^{15}$N and $^1$H-$^{13}$C skyline projection in reconstructed 3D HNCO spectra [21] from 30% NUS data. (a) denotes the fully sampled spectrum, (b) and (c) are reconstruction spectra at the time of 1.32h by LRHM and LRHMF, respectively. (d) represents the reconstruction at the time of 27.8h by LRHM. Note: The mentioned time is based on the reconstruction algorithms without parallel acceleration. Adapted from Ref. [11b].

As indicated in Figure 6, under NUS, the reconstruction can be carried out independently. Parallel computation is a technology that makes it possible to reconstruct several indirect planes simultaneously, which accelerates the reconstruction, especially when the reconstruction in each indirect plane takes time.

We apply the parallel computation in 2D and 3D spectra reconstruction. The setting of the parallel computing environment and the computation time has been recorded in Table 1.

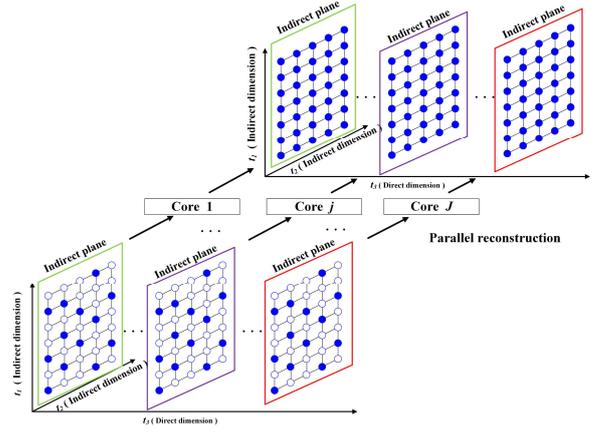

**Figure 11.** The sketch map of parallel computation of 3D spectra reconstruction. Adapted from Ref. [11b].

Compared with LRHM, LRHMF approximately decreases 67 times of computational time in 3D spectra reconstruction.

In some applications, such as the spatiotemporally encoded ultrafast (STEU) NMR spectroscopy, NUS is applied to save the sampling time. Unlike the standard NMR spectrum, the 2D STEU spectrum is obtained in a hybrid time and frequency (HTF) plane, involving the reconstruction of NUS within the mixing frequency domain. It has been proved that LRHMF can be used to solve this problem [11a], which shows the advantages of leveraging the low rank Hankel property. This method, called LRBHM-HTF, can be expressed as

$$\min_{P,Q,G} \frac{1}{2}\left(\|\mathbf{P}\|_F^2 + \|\mathbf{Q}\|_F^2\right) + \frac{\lambda}{2}\|\mathbf{Y} - \mathcal{U}\mathbf{G}\|_F^2 \quad s.t. \mathcal{B}\mathcal{F}_{freq}^{-1}\mathbf{G} = \mathbf{P}\mathbf{Q}^H, \quad (19)$$

where $\mathbf{G}$ represents a two-dimensional HTF signal, and $\mathcal{F}_{freq}^{-1}$ is an operator to transform every row vector of HTF with 1D discrete Fourier transform into a time-domain signal.

This method is verified on a real 2D STEU COSY of a special type of oil existing in intrahepatic fat of the liver [11a]. Results in Figure 12 illustrate that, compared with CS, LRBHM-HTF enhances peak intensities and suppresses fake peaks, nearly restoring the information of the original spectrum, even when the NUS reconstruction problem is extended to the HTF plane.

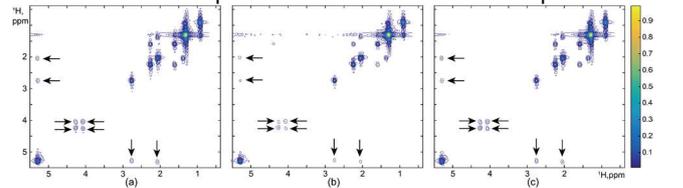

**Figure 12.** Reconstruction of STEU COSY spectrum. (a) The full sampled spectrum; (b) and (c) The reconstructed spectrum by CS and LRBHM-HTF method from 20% NUS data, respectively. Adapted from Ref. [11a].

We evaluate the reconstruction by comparing the proton resonance volume of two methods. The comparison in Table 7 indicates that the resonance volume of the reconstruction by LRBHM-HTF is much more robust and closer to the reference than CS.

Furthermore, in order to better analyze the results in Table 7, we define the Quantitative Error Factor (QEF) as [11a]

$$\text{QEF} = \frac{|s - \hat{s}|}{s}, \quad (20)$$

where $s$ and $\hat{s}$ denote the resonance volume of the reference and the reconstructions, respectively.

The results shown in Figure 13 confirm our claims in Table 7. LRBHM-HTF owns lower QEF and smaller error bars than CS. Moreover, for most of the peaks with high resonance volume (such as Peak 2, 3, and 4), QEF of both methods is sufficiently low, implies that both two methods achieve nice results. For low-resonance peaks (such as Peak 5-9), QEF of CS is much higher than LRBHM-HTF. For Peak 1, with the largest resonance volume in the spectrum, CS enhanced its intensity too much while LRBHM-HTF provided a more reasonable result [11a].

**Table 7.** The proton resonance volume comparison between CS and LRBHM-HTF [11a].

| Protons | | | Resonance volume | | |
|---|---|---|---|---|---|
| ID | Type | ppm | Reference | CS | LRBHM-HTF |
| 1 | methylene | 1.30 | 37.69 | $40.30 \pm 0.13$ | $38.01 \pm 0.03$ |
| 2 | methyl | 0.90 | 16.79 | $16.73 \pm 0.15$ | $16.80 \pm 0.02$ |
| 3 | olefinic | 5.29 | 12.98 | $12.92 \pm 0.20$ | $13.00 \pm 0.01$ |
| 4 | $\alpha$-olefinic | 2.02 | 12.08 | $11.79 \pm 0.16$ | $12.06 \pm 0.03$ |
| 5 | $\alpha$-carboxyl | 2.24 | 8.16 | $7.69 \pm 0.18$ | $8.12 \pm 0.04$ |
| 6 | diacyl methylene | 2.75 | 5.23 | $4.64 \pm 0.11$ | $5.17 \pm 0.03$ |
| 7 | $\beta$-carboxyl | 1.60 | 4.78 | $4.28 \pm 0.18$ | $4.63 \pm 0.02$ |
| 8 | glecerol backbone CH | 4.30 | 1.19 | $0.86 \pm 0.04$ | $1.16 \pm 0.01$ |
| 9 | $CH_2$ | 4.05 | 1.10 | $0.79 \pm 0.02$ | $1.05 \pm 0.01$ |

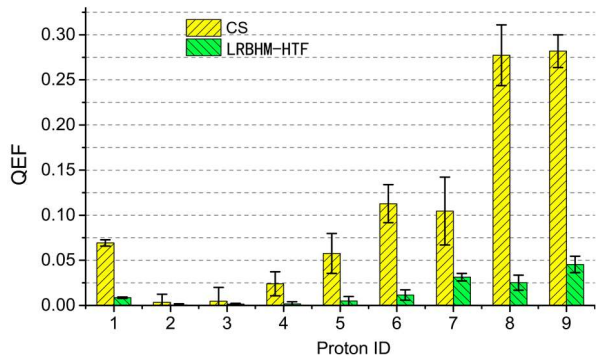

**Figure 13.** QEF of the proton resonance volume in Table 7. Note: the vertical bar comes from the randomness of sampling schedules [11a]. Adapted from Ref. [11a].

### 3.3. N-D Spectra Reconstruction with Low Rank Tensor

In theory, for $N$-dimensional ($N \geq 3$) spectra, spectral signals can be recovered by minimizing the rank of $N$-fold Hankel matrix [6b], which theoretically ensures the stability of recovery. This approach nevertheless invokes the problem of minimizing a matrix with too large size [6c] and prohibits its application to high dimensional signals.

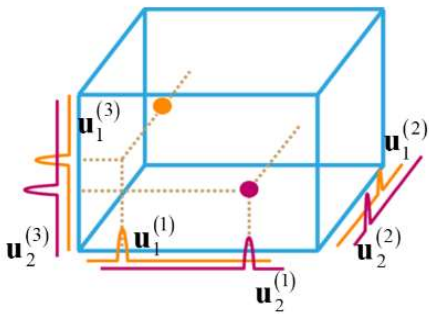

**Figure 14.** CANDECOMP/PARAFAC (CP) tensor decomposition for 3D exponential signals.

Another idea is to reduce the dimension of $N$-D signals, transforming this issue into the reconstruction in a lower dimension. Therefore, the decomposition of tensor was taken into consideration. Taking a 3D case as an example (Figure 14), the signal can be treated as a cube with CANDECOMP/PARAFAC (CP) tensor. The small number of peaks means that the signal is "low-CP-rank". The typical reconstruction method, Multi-Dimensional Decomposition (MDD) [22], assumes that N-D spectral signals own a low-CP-rank structure. This property has also been utilized in a typical non-convex approach [23], named as the CP-Tensor method, in matrix completion issue. Ying et al. [6c] proposed a method, called Hankel Matrix nuclear norm Regularized low-CP-rank Tensor Completion (HMRTC), to simultaneously exploits the low-CP-rank structure and the exponential structure via constraining the low rank Hankel property of the associated factor vectors. This method is expressed as

$$\min_{\substack{\mathbf{u}_r^{(n)} \\ r=1,\ldots,R \\ n=1,\ldots,N}} \sum_{r=1}^{R}\sum_{n=1}^{N} \left\| \mathcal{R}\mathbf{u}_r^{(n)} \right\|_* + \frac{\lambda}{2}\left\| \mathcal{Y} - \mathcal{U}\mathcal{X} \right\|_F^2, \qquad (21)$$

where $\mathcal{X} = \sum_{r=1}^{R} \mathbf{u}_r^{(1)} \otimes \mathbf{u}_r^{(2)} \otimes \cdots \otimes \mathbf{u}_r^{(N)} \in \mathbb{C}^{I_1 \times I_2 \times I_N}$ are N-D signals with $R$ components in the time domain, $\otimes$ is the outer product, and $\mathcal{Y}$ is the NUS data with noise.

The solver of HMRTC and its convergence has been provided [6c]. HMRTC is validated on a 3D HNCO spectrum and compared with CP-Tensor in Figure 15. Under a very limited sampling rate (10%), HMRTC still offers a high-quality spectrum while CP-Tensor weakens peak intensity and loses a number of peaks.

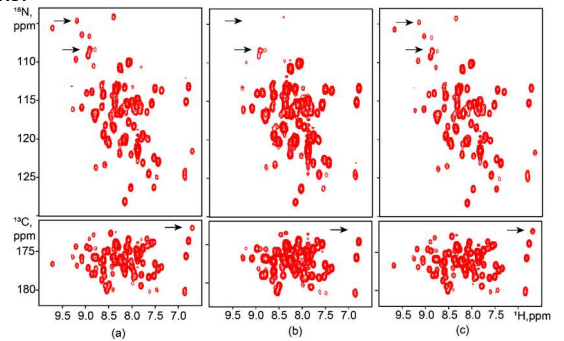

**Figure 15.** The reconstructed spectra of a 3D HNCO experiment[21]. (a) denotes the fully sampled spectrum; (b) and (c) are the CP-Tensor and HMRTC reconstructions with 10% 3D NUS data, respectively. The parameter setting of CP-Tensor and HMRTC is explained in [6c]. Adapted from Ref. [6c].

### 3.4. Conclusion on Typical Low Rank Hankel Matrix/Tensor Reconstruction

Typical low rank Hankel matrix/tensor methods exploit the intrinsic property of exponential and is considered to be theoretically best adapted to NMR signals [19], enabling to overcome some difficulties that are troublesome to other conventional methods [6a, 20], such as the reconstruction of broad peaks.

However, their drawbacks are also obvious. The computation time of these methods remains an issue, especially for high-dimensional or large number point signals. Nevertheless, some new algorithms are proposed to decrease a large amount of computational time.

In some methods of reconstruction, the nuclear norm is the surrogate to the rank. For SVD, both $\Sigma$ and two unitary matrices ($\mathbf{U}$ and $\mathbf{V}$) can be exploited in the reconstruction issue. LRHM minimizes the rank of the Hankel matrix given by the FID signal (i.e., the number of peaks). The adopted nuclear norm is a type of convex approximation of the rank so that some convex algorithms, such as ADMM[6b] and SVT[27], can be employed to solve the low rank reconstruction problem. However, compared with the original problem, this convex approximation requires more NUS data points[28]. Besides, as is mentioned in Figure 1, the subspace set $\left\{\mathbf{x}_r = \mathcal{R}^* \mathbf{u}_r \mathbf{v}_r^H \right\}_{r=1,2,\cdots,R}$, which is derived from matrices $\mathbf{U}$ and $\mathbf{V}$, contains the line shape information of peaks, which is ignored in LRHM.

## 4. Enhanced Low Rank Hankel Matrix Reconstruction

### 4.1. Better Approximation to the Rank

As stated above, in typical low rank Hankel matrix/tensor reconstruction methods, such as LRHM, although the convex relaxation of the rank avoids the NP-hard, another problem emerges that the relaxation puts weights on the singular values (Figure 16(a)). This problem results in imperfect performance in the reconstruction with a low sampling rate or low-intensity peaks (Figure 16(b) and (c)).

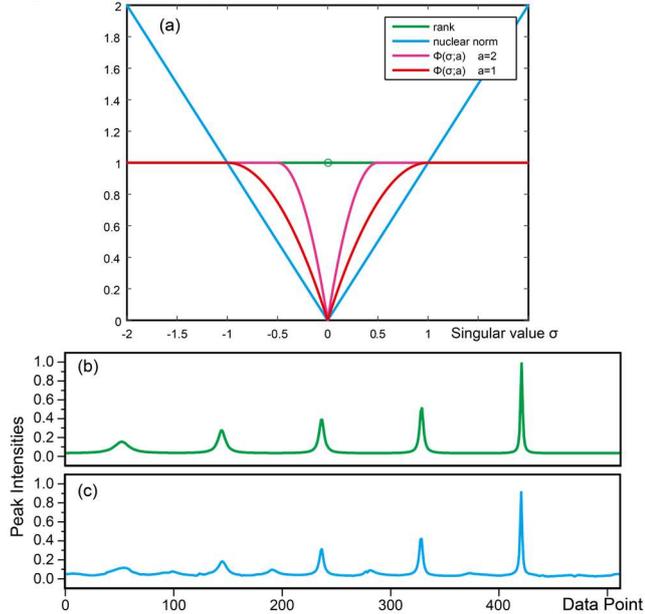

**Figure 16.** The limitations of LRHM. (a) Comparison between rank function and nuclear norm. (b) and (c) are the fully sampled signal and the reconstruction result with 10% sampled data, respectively. Adapted from Ref. [13].

Two enhanced low rank Hankel methods are proposed based on replacing the nuclear norm by other constraints. One is Weighted Low Rank Hankel matrix (WLR) [12], the other is Enhanced Low Rank Hankel matrix (ECLR) [13].

In WLR, a weighted nuclear norm has been introduced to substitute the nuclear norm, and its definition is presented as

$$\|\mathbf{X}\|_{\mathbf{w},*} = \sum_{q=1}^{Q} w_q \sigma_q, \qquad (22)$$

where $\mathbf{w} = \begin{bmatrix} w_1 & \cdots & w_q & \cdots & w_Q \end{bmatrix}^T$ is the weight, and $w_q$ denotes the weight put on the $q^{th}$ singular value $\sigma_q$.

The WLR method is to solve such a problem:

$$\min_{\mathbf{x}} \|\mathcal{R}\mathbf{x}\|_{\mathbf{w},*} + \frac{\lambda}{2} \|\mathbf{y} - \mathcal{U}\mathbf{x}\|_2^2. \qquad (23)$$

According to Figure 16(a), ideal weights should be located at the reciprocal of non-zero singular values so that the constraint approximates the rank function. In practice, however, neither the true rank nor singular values are known. Thus, we use LRHM to obtain pre-reconstruction, utilize its rank and singular values to replace real ones, and update them in each outer loop.

**Table 8.** The solver of WLR.

**Initialization**: Input $\mathbf{y}$, $\lambda$, $\mathcal{U}$, set $P = 4$, $k_{max} = 10^3$.

Initialize $\mathbf{x}_0 = \mathcal{U}^* \mathbf{y}$, $\mathbf{D}_0 = \mathbf{1}$, $k = 0$, $\dfrac{\|\mathbf{x}_{k+1} - \mathbf{x}_k\|_2}{\|\mathbf{x}_k\|_2} = 1$, and obtain the pre-reconstruction $\tilde{\mathbf{x}}$ from LRHM.

**Main:**
**For** each time $p = 1, 2, \cdots, P$ of updating weights, **do:**
1) Estimate $\tilde{\mathbf{P}}$ from $\tilde{\mathbf{x}}$ according to $\mathcal{R}\mathbf{x} = \tilde{\mathbf{U}} \tilde{\Sigma} \tilde{\mathbf{V}}^H$;
2) Compute $\mathbf{w} = \begin{bmatrix} w_1 & \cdots & w_q & \cdots & w_Q \end{bmatrix}^T$ according to $w_q = \left( \tilde{\Sigma}_q + \upsilon \right)^{-1}$

    **While** $k < k_{max}$ and $\dfrac{\|\mathbf{x}_{k+1} - \mathbf{x}_k\|_2}{\|\mathbf{x}_k\|_2} \leq 10^{-6}$, **do**
      a) Update
$$\mathbf{x}_{k+1} = \left( \beta \mathcal{R}^* \mathcal{R} + \lambda \mathcal{U}^* \mathcal{U} \right)^{-1} \left( \lambda \mathcal{U}^* \mathbf{y} + \beta \mathcal{R}^* \left( \mathbf{Z}_k - \mathbf{D}_k / \beta \right) \right);$$
      b) Update $\tilde{\mathbf{U}}_{k+1} \Sigma_{k+1} \mathbf{V}_{k+1}^H = \mathcal{R}\mathbf{x}_{k+1} + \mathbf{D}_k / \beta$;
      c) Update
$$\mathbf{Z}_{k+1} = \tilde{\mathbf{U}}_{k+1} \left( \Sigma_{k+1} - diag(\mathbf{w}) / \beta \right) \mathbf{V}_{k+1}^H;$$
      d) Update $\mathbf{D}_{k+1} = \mathbf{D}_k + (\mathcal{R}\mathbf{x}_{k+1} - \mathbf{Z}_{k+1})$;
      e) $k \leftarrow k + 1$
    **End while**
3) $k = 0$, $\dfrac{\|\mathbf{x}_{k+1} - \mathbf{x}_k\|_2}{\|\mathbf{x}_k\|_2} = 1$, $p \leftarrow p + 1$, $\tilde{\mathbf{x}} \leftarrow \mathbf{x}_{k+1}$;

**End while**
**Output:** $\tilde{\mathbf{x}}$.

And for ECLR, it is stated as

$$\min_{\mathbf{x}} \sum_{q=1}^{Q} \phi \left( \sigma_q (\mathcal{R}\mathbf{x}); a \right) + \frac{\lambda}{2} \|\mathbf{y} - \mathcal{U}\mathbf{x}\|_2^2, \qquad (24)$$

where $Q$ stands for the total number of the singular values in Hankel matrix. The function $\phi$ is given by

$$\phi \left( \sigma_q (\mathcal{R}\mathbf{x}); a \right) = \begin{cases} |\sigma_q| - \dfrac{a}{2} \sigma_q^2, & |\sigma_q| \leq \dfrac{1}{a} \\ \dfrac{1}{2a}, & |\sigma_q| \leq \dfrac{1}{a} \end{cases}, \qquad (25)$$

where $\sigma_q$ stands for the $q^{th}$ singular value of the matrix, and $a$ is a parameter which controls the distance of the value of the function $\phi$ to the actual rank (The red and pink lines in Figure

16(a)).

In Table 9, $\Theta$ denotes a thresholding function on singular values stored in the diagonal entry in $\Sigma$ and is defined as

$$\Theta(\Sigma;\beta,a) = \min\left\{\Sigma, \max\left\{(\Sigma - 2a/\beta)/(1 - 2a^2/\beta), 0\right\}\right\}. \quad (26)$$

**Table 9.** The solver of ECLR.

**Initialization:** Input $\mathbf{y}$, $\mathcal{R}$, $\lambda$, $\mathcal{U}$, set $j=0$, $k=0$, $a=0$, $b=0.2$, $\beta=1$, $k_{\max}=500$, $\beta_{\max}=2^{20}$ and $\mathbf{x}_0=\mathbf{y}$.

**Main:**

While $\beta_j < \beta_{\max}$ and $\dfrac{\|\mathbf{x}_{j+1} - \mathbf{x}_j\|_2}{\|\mathbf{x}_j\|_2} \leq 10^{-4}$, do:

   While $k < k_{\max}$ and $\dfrac{\|\mathbf{x}_{k+1} - \mathbf{x}_k\|_2}{\|\mathbf{x}_k\|_2} \leq 10^{-4}$, do:

   1) Update $\mathcal{R}\mathbf{x}_{k+1} = \mathbf{U}_{k+1}\Sigma_{k+1}\mathbf{V}_{k+1}^H$;

   2) Update $\mathbf{Z}_{k+1} = \mathbf{U}_{k+1}\Theta\left(\Sigma_{k+1}; \dfrac{2a}{\beta_j}, a\right)\mathbf{V}_{k+1}^H$;

   3) Update $\mathbf{x}_{k+1} = \left(\beta_j \mathcal{R}^*\mathcal{R} + \lambda \mathcal{U}^*\mathcal{U}\right)^{-1}\left(\beta_j \mathcal{R}^*\mathbf{Z}_k + \lambda \mathcal{U}^*\mathbf{y}\right)$;

   4) $k \leftarrow k+1$

   **End while**

   5) $k=0$, $\beta \leftarrow 2\beta$, $a = \sqrt{b\beta/2}$, $\mathbf{x}_{j+1} \leftarrow \mathbf{x}_{k+1}$;

**End while**

**Output:** $\mathbf{x}$.

### 4.2. Subspace Constraint via Vandermonde Factorization

LRHM and LRHMF exploit the exponential property of FID signals via minimizing the number of non-zero singular values of the Hankel/block Hankel matrix given by FID signals. However, as is mentioned above, the spectral line shape information (including the amplitude, location and width of peaks) in the subspace set $\{\mathbf{x}_r = \mathcal{R}^*\mathbf{u}_r\mathbf{v}_r^H\}_{r=1,2,\cdots,R}$ is ignored. Furthermore, the SVD is not the most suitable way to decompose exponentials, because the subspace set $\{\mathbf{x}_r = \mathcal{R}^*\mathbf{u}_r\mathbf{v}_r^H\}_{r=1,2,\cdots,R}$ does not correspond to the actual exponentials and their physical meaning is still unclear. Introducing a more proper decomposition way and exploiting the subspaces of this new decomposition may promote the fidelity of reconstructed spectra.

A method called Hankel matrix completed by Vandermonde factorization (HVaF) [14] is available. Unlike SVD, Vandermonde factorization of the Hankel matrix generates a series of exponential subspaces, which correspond to the exponential components of FID signals in theory. HVaF constrains the low rank Hankel properties of these exponential subspaces, allowing better use of the exponential structure of FID signals than LRHM.

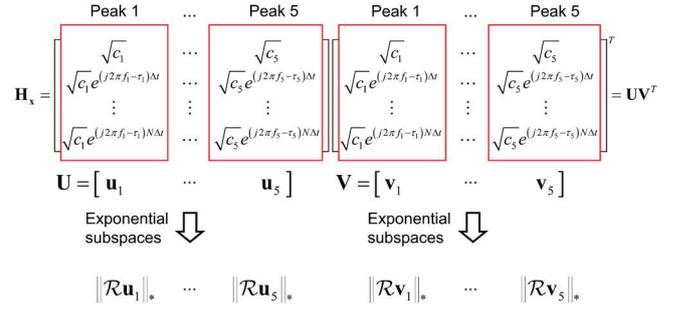

**Figure 17.** The key idea of HVaF.

HVaF is written as

$$\min_{\mathbf{U},\mathbf{V},\mathbf{x}} \sum_{r=1}^{\hat{R}} \left(\|\mathcal{R}_m\mathcal{Q}_r\mathbf{U}\|_* + \|\mathcal{R}_m\mathcal{Q}_r\mathbf{V}\|_*\right) + \frac{\lambda}{2}\|\mathcal{U}\mathbf{x} - \mathbf{y}\|_2^2, \quad \text{s.t. } \mathcal{R}\mathbf{x} = \mathbf{U}\mathbf{V}^T, \quad (27)$$

where $\mathcal{Q}_r : \mathbb{C}^{M \times N} \to \mathbb{C}^{M \times 1}$ represents the operation of extracting $r^{th}$ column of the matrix.

Since the factorization in HVaF is theoretically adapted to NMR spectral signals, this method is expected to provide an accurate estimate of parameters, including the amplitude, the central frequency, and the damped factor [14].

### 4.3. Conclusion on Enhanced Low Rank Hankel Matrix Reconstruction

Here, we analyze the reconstruction performances of LRHM, WLR, ECLR, and HVaF on 1D synthetic data and 2D realistic NMR data. The 2D realistic NMR data is from a $^1$H-$^{15}$N HSQC spectrum of intrinsically disordered cytosolic domain of human CD79b protein from the B-cell receptor.

Firstly, we compare WLR, ECLR, and HVaF with LRHM on the 2D HSQC spectrum (Figure 18). All four methods still achieve reasonable results under a high acceleration factor (5 factors). LRHM yields a spectrum with several peaks diminished and lost. Compared with LRHM, WLR, ECLR, and HVaF offer better results with a higher Pearson coefficient $R^2$, promoting peak intensity, especially low intensity (lower than $2\times10^7$). For WLR, some artifacts do appear while ECLR and HVaF suppress artifacts better.

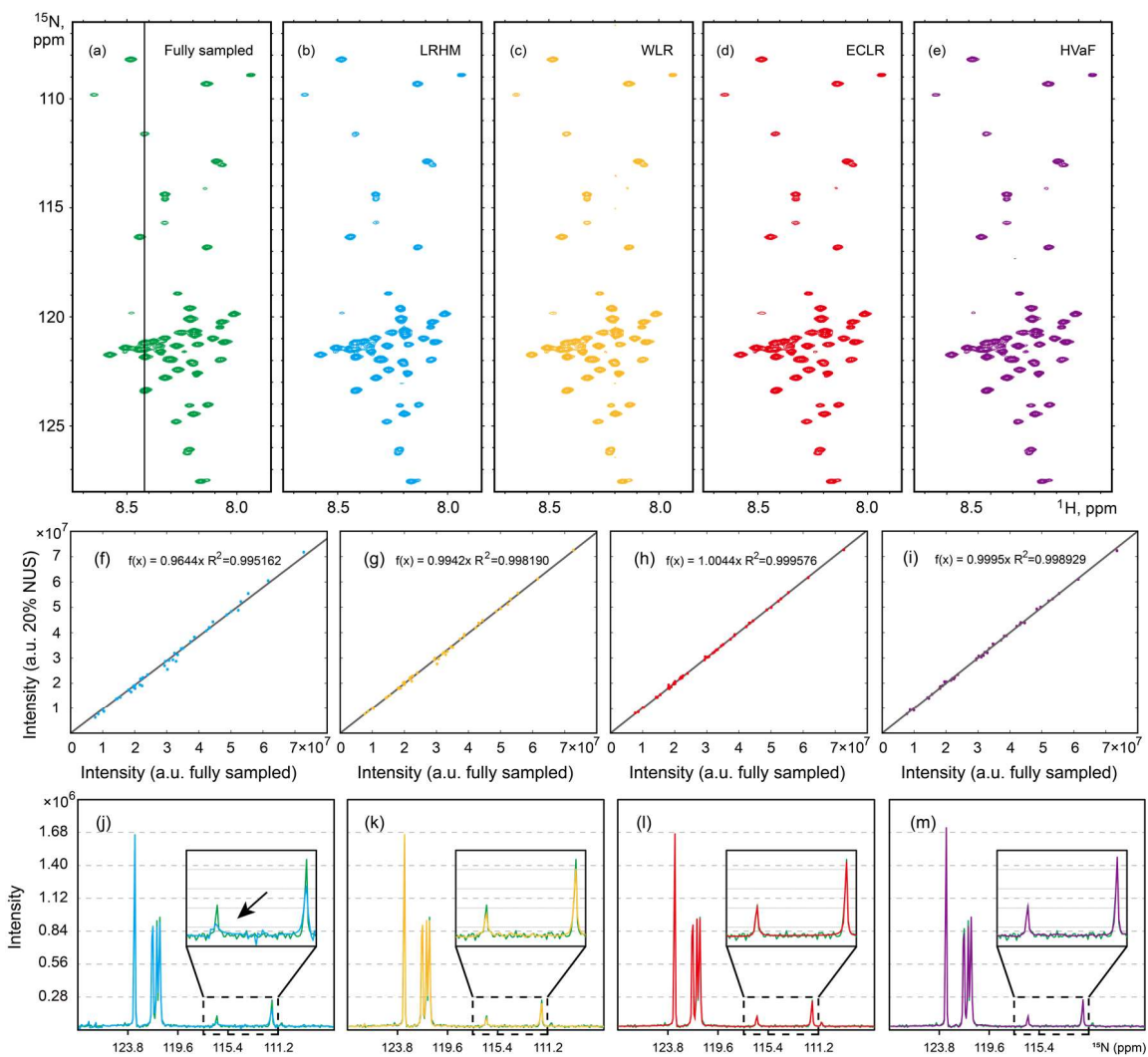

**Figure 18.** The reconstructions of the 2D HSQC spectrum from 20% NUS data. (a) denotes the fully sampled spectrum. (b)-(e) are reconstructed spectra of LRHM, WLR, ECLR, and HVaF, respectively. (f)-(i) stand for the peak intensity correlation for four methods. In (j)-(m), green, blue, yellow, red, and purple lines stand for the 1D $^{15}$N trace at 8.42 ppm (the black line in (a)) in the full sampling, LRHM, WLR, ECLR, and HVaF, respectively.

To better evaluate the performance of these four methods, we test them at different sampling rates (8%, 10%, and 15%) on 1D synthetic data (Figure 19).

At a relatively high NUS level (15%), four methods generate reasonable results, although LRHM slightly distorts the shape peaks of low intensity (the first two peaks). As the decreasing of the NUS level (10% and 8%), the reconstructed spectra of WLR, ECLR, and HVaF remain the high-fidelity, LRHM, however, severely distorts and weakens peaks (marked by black arrows).

We utilize the peak intensities correlation of reconstructions in 100 Monte Carlo trials to evaluate the line shape of peaks. The average and the standard deviation are recorded in Figure 19, showing that these three enhanced methods offer higher and more robust correlation values compared with LRHM, especially for peaks with low intensity. Taking Peak 1 as an example, at these three NUS levels, three enhanced methods improve the correlation, providing higher averages and relatively smaller standard deviations. Furthermore, compared with WLR, ECLR and HVaF generate more accurate and robust correlations. However, under very limited NUS measurement (8%), for some peaks (such as Peak 3), ECLR and HVaF produce worse correlation compared with LRHM.

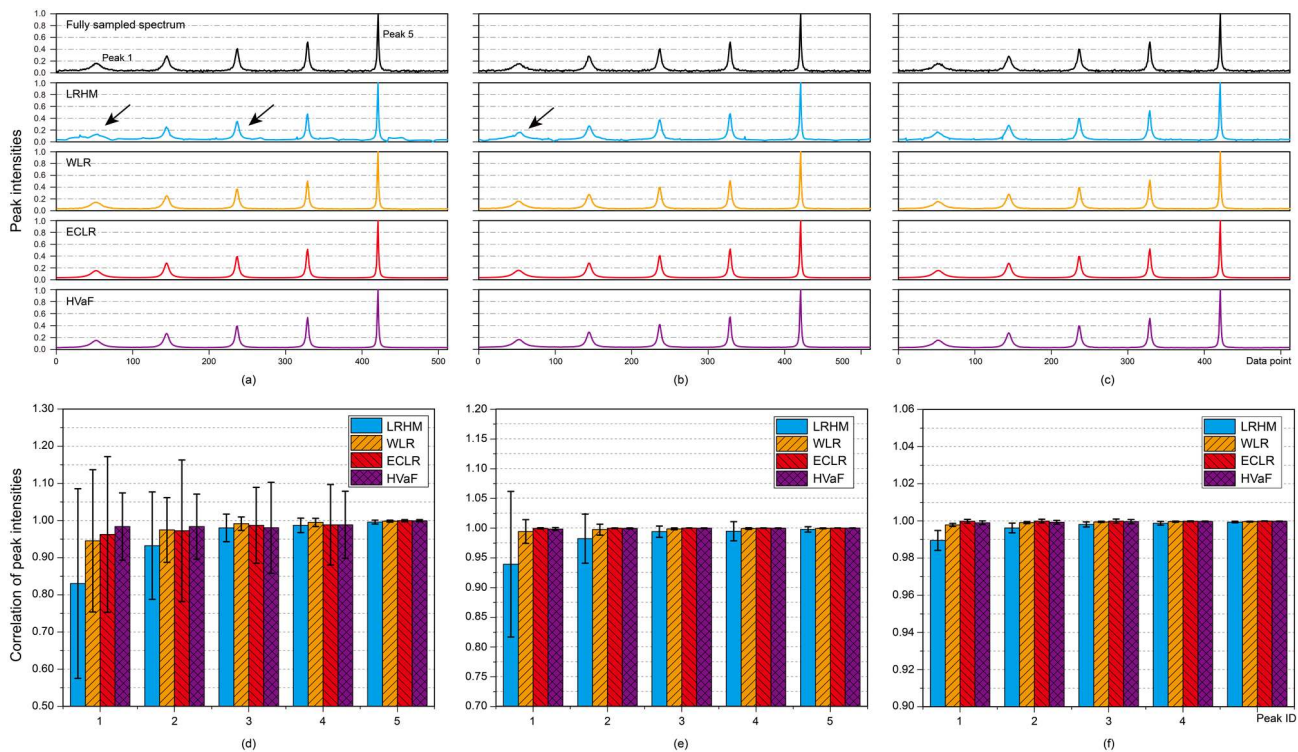

**Figure 19.** The reconstructions of 1D synthetic spectrum from 8%, 10% and 15% NUS data. (a)-(c) denote fully sampled spectra with noise ($\sigma = 0.005$) and reconstructed spectra of LRHM, WLR, ECLR and HVaF from 8%, 10% and 15% NUS data. (d)-(f) are the peak intensity correlation for four methods. The vertical bars come from the randomness of NUS schedules in 100 Monte Carlo trials.

The joint distribution of correlation for Peak 3 is presented in Figure 20 to compare the performance of four methods in each trial. Taking Figure 20(a) as an example, in each trial, the correlation of WLR and LRHM are denoted as the horizontal axis value and the vertical axis value of one point, respectively. WLR is claimed to achieve a higher correlation than LRHM if this point is located above the dash-dot line.

For Peak 3, in most of the trials, compared with LRHM, enhanced methods obtain relatively higher correlation values, meaning that peaks of the reconstructed spectra are closer to those of the fully sampled spectrum. However, in several trials, ECLR and HVaF offer some extremely low correlation (marked by arrows in Figure 20(b) and (c)). These abnormal values result in lower average and larger standard deviation.

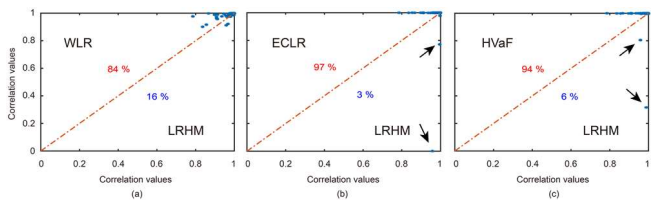

**Figure 20.** The joint distribution of correlation for Peak 3. (a)-(c) denote the joint distribution of LRHM, WLR, ECLR, and HVaF from 8% NUS data, respectively. Note: For each Monte Carlo trial, the method presented in the vertical axis, denoted as method A, is claimed to improve the correlation than the method presented in the horizontal axis, denoted as method B, if the point lies above the dash-dot line. The red (or blue) number stands for the percentage that method A reaches a higher (or lower) correlation than method B.

In summary, results on synthetic and experimental data confirm that enhanced methods do promote the fidelity of reconstructed spectra. These three enhanced methods improve the basic LRHM from different aspects: 1) WLR and ECLR introduce new approximations that are closer to the rank function than the nuclear norm; 2) HVaF exploits both exponential subspaces and the number of non-zero singular values. Nevertheless, enhanced methods might be trapped into local optimum since the numerical algorithms are non-convex. Besides, enhanced methods cost a longer computational time than LRHM. All in all, how to fast reconstruct high-quality spectra from NUS data is still a problem worthy of in-depth study.

## 5. Summary and Outlook

In this review, we demonstrate that low rank Hankel matrix/tensor methods play an essential role in denoising and reconstructing NMR spectra. So far, non-uniformly sampling has been an alternative approach to overcome the long acquisition time and low SNR problems. The FID signals can be modeled as the sum of complex exponentials of which Hankel matrices are often low rank. These low rank Hankel matrix/tensor methods have shown good performance in denoising low SNR spectra and reconstructing non-uniformly sampled spectra under high acceleration factors. The long computational time and the existence of non-exponential signals, however, for these reconstruction methods limit their wider spread.

Deep Learning (DL) is a representative artificial intelligence technique utilizing neural networks [29], which has the advantages of high-speed, end to end, and user-friendly. Recently, it has been successfully used for the fast reconstruction of high-quality NMR spectra [9, 24]. These works prove that the neural network of DL can be successfully trained using solely synthetic NMR data with exponential functions [6c, 11a, 12, 14]. We think that introducing the low rank Hankel property of exponential functions may improve the stability of DL, enhance its interpretability, and make it a much more efficient and powerful technique in NMR spectroscopy [30].

With the future development of low rank Hankel matrix/tensor methods, we may anticipate that more problems in NMR spectroscopy will be solved. An incomplete list of possible future wok may include: (1) Develop faster algorithms with new numerical solvers or high-performance computing hardware. (2) Reconstruct higher-dimensional NMR and functional NMR, such as diffusion and time-resolved spectra. (3) Combine the exponential structure and Hankel matrices/tensors with artificial intelligence to enable more accurate and faster reconstruction.

## Acknowledgements

Xiaobo Qu thanks Prof. Zhi-Pei Liang, from University of Illinois at Urbana-Champaign, for the generous invitation to join Prof. Liang's lab in 2009-2011. During the visit, the low rank Hankel matrix property of exponential functions[31] came into his horizon and inspired him to complete a series of low rank Hankel reconstruction methods in NMR spectroscopy. The authors would like to thank Prof. Zhong Chen from Xiamen University, and Prof. Vladislav Orekhov from University of Gothenburg, for valuable discussions in collaboration. The authors thank Wanqi Hu, Dongbao Liu, Xianfeng Chen and Khan Asfar for preparing some figures and polishing English. The authors are also deeply grateful to other researchers for insightful discussions and publishers for permission to adopt figures. Besides, the authors thank editors and reviewers for valuable suggestions. This work was supported in part by the National Natural Science Foundation of China (NSFC) under grants 61971361, 61871341 and U1632274, the Joint NSFC-Swedish Foundation for International Cooperation in Research and Higher Education (STINT) under grant 61811530021, the National Key R&D Program of China under grant 2017YFC0108703, the Natural Science Foundation of Fujian Province of China under grant 2018J06018, the Fundamental Research Funds for the Central Universities under grant 20720180056, and the Xiamen University Nanqiang Outstanding Talents Program.